\documentclass[epj,nopacs]{svjour}
\usepackage{graphics}
\usepackage{subfigure}
\usepackage{bm}

\begin{document}
\title{Elliptic flow in proton-proton collisions at $\sqrt{s}=7$~TeV}
\author{Piotr Bo\.zek\inst{1}\inst{2}
}                     
%
%
\institute{The H. Niewodnicza\'nski Institute of Nuclear Physics,
PL-31342 Krak\'ow, Poland, \email{piotr.bozek@ifj.edu.pl} \and
Institute of Physics, Rzesz\'ow University, 
PL-35959 Rzesz\'ow, Poland}
\date{Received: date / Revised version: date}
%
\abstract{The angular correlations measured in proton-proton collisions at
 $\sqrt{s}=7$~TeV are decomposed into contributions from 
back to back emission  and elliptic 
flow.  Modeling the dominant term in the correlation functions 
as a momentum conservation effect or as an effect of the initial
 transverse velocity of the source, the remaining elliptic flow
 component can be estimated. 
The elliptic flow coefficient extracted from the CMS Collaboration data
 is $0.04-0.08$.  No additional
 small-angle, 
ridge-like correlations are needed to explain the experimental data.
} 
\maketitle
\section{Introduction}

Proton-proton collisions at the CERN Large Hadron Collider with 
the energy of $7$~TeV are the most violent elementary collisions studied in a
laboratory. 
The high multiplicity of produced particles means that a
short-lived, high density system is formed. By analogy to the physics
of ultrarelativistic heavy-ion collisions, one could expect new 
collective phenomena to emerge in the highest multiplicity events.
A striking experimental observation has been made by the CMS Collaboration,
measuring two-particle correlations $R(\Delta \phi, \Delta \eta)$
 in the relative azimuthal angle and the relative pseudorapidity 
\cite{Khachatryan:2010gv}.  In high multiplicity events an
 enhancement of the correlation on the near side ($\Delta \phi \simeq 0$) 
is observed, extending in pseudorapidity in the form of  a ridge. 
This gives rise to speculations on the origin of the effect 
\cite{Shuryak:2010wp,Dumitru:2010iy}.

We present a quantitative analysis of the angular correlation $R(\Delta \phi)$
 measured by the CMS experiment for pair of particles separated in pseudorapidity $|\Delta \eta|>2.0$. 
The angular correlation function is 
decomposed into a contribution from global momentum flow and
 a smaller elliptic component. 
The dominant structure in the correlations function in the 
relative azimuthal angle is due to an enhancement of back to back emission.
This effect
is qualitatively described by  Monte Carlo event generators
 \cite{Khachatryan:2010gv}. The experimental data compared to the 
PYTHIA generator results show two essential differences. For 
all transverse momentum and multiplicity classes presented, the measured
angular correlations show deviations from model calculations. In simulations
 and in the data, 
the
 correlations are enhanced for the 
 emission with $\Delta \phi \simeq \pi$, but the 
width and the height of the peak is different. The most important new 
feature in the measured  correlation functions is the appearance 
of a small-angle enhancement of the distribution. This small-angle 
structure (ridge) is 
present in a broad range in the  relative pseudorapidity between 
 the particles
 in the pair. This small-angle enhancement is not due to 
the usual correlations from the jet fragmentation, and is 
not reproduced by any Monte Carlo event generator. 

We analyze
 the possibility that the new effect is due to the elliptic flow present
besides the dominant back to back angular correlations. For some transverse
 momentum and multiplicity cuts, the relative importance of the elliptic
 flow component is big enough to make it explicitly 
visible  in the small-angle region. However, the 
 second harmonic $\cos(2\Delta \phi) $  
cannot be unambiguously unraveled in the whole range ($0$-$\pi$)
from the dominant background in the correlation function.

The importance of the elliptic flow is that it is 
 a signature of the collective expansion of 
the fireball created in relativistic heavy-ion collisions
\cite{Ollitrault:1992}. It is important  to
 check whether similar phenomena occur in elementary collisions, but
 the search for the elliptic flow in
 smaller systems  is  hindered by statistical
 fluctuations and non-flow correlations. Several model 
estimates of the elliptic flow generated in proton-proton 
collisions have been given  \cite{Cunqueiro:2008uu,Luzum:2009sb,Bautista:2009my,d'Enterria:2010hd,Prasad:2009bx,Bozek:2009dt,CasalderreySolana:2009uk,Ortona:2009yc,Avsar:2010rf}.
 Most commonly, one assumes that the 
elliptic flow is generated during a short hydrodynamic expansion, 
in a very similar way as in heavy-ion
 collisions. The most important differences between calculations
 concern the origin of the azimuthal asymmetry of the
 initial energy density distribution in the transverse plane.
 If the presence of an elliptic flow of collective origin could 
be demonstrated in proton-proton collisions, it would indicate
 the a strongly interacting fireball has been formed in hadron 
collisions at the highest energies.

The elliptic flow correlations are subleading. The extraction of the
 elliptic flow coefficient requires a careful analysis of other effects.
In the following, we propose two models for the dominant back to back
 correlations. The first estimate is based on the momentum conservation 
effects, requiring the conservation of the total transverse momentum. The 
second calculation assumes that particles separated in pseudorapidity are
 emitted from sources moving (on average) in opposite  
transverse directions. This picture is natural if the two particles come
 from two different back to back jets. In a high multiplicity event, and 
for low transverse momenta of the particles, a simple picture can be used, 
with 
two  sources moving with opposite transverse velocities.

To reproduce the observed ridge correlations, an additional azimuthal 
asymmetric component (elliptic flow) is added.
The estimated elliptic flow coefficient is $v_2=0.06$-$0.1$, 
for the parameters where the ridge is observed.
The elliptic flow interpretation agrees qualitatively with the multiplicity
 and transverse momentum dependence of the observed effect. 
 The strength of the small-angle 
two-particle correlations is proportional to $v_2^2$,
 and should increase as $p_\perp^2$.

\section{Momentum conservation}

The correlation function $R(\Delta \phi)$ is obtained selecting  
particles separated in pseudorapidity \cite{Khachatryan:2010gv}. 
This procedure eliminates
jet-like correlations on the near side. The form of the 
dominant remaining correlations 
shows, that  the associated particle is emitted preferentially 
in the opposite direction in the azimuthal angle. This could be  a consequence 
of the momentum conservation in the microscopic particle production mechanism.
Such correlations can be very important in small multiplicity events, 
as in proton-proton collisions \cite{Borghini:2000cm,Borghini:2002mv,Chajecki:2008yi,Bzdak:2010fd}.
In the following we take into account the  transverse momentum
 conservation in the same way as proposed in \cite{Borghini:2000cm,Bzdak:2010fd}.
In an event of multiplicity $M$, the multiparticle distribution in transverse
momenta $\vec{p}_i$ and pseudorapidities $\eta_i$ 
is given
as  

\begin{figure*}[ht]
\resizebox{0.75\textwidth}{!}{%
  \includegraphics{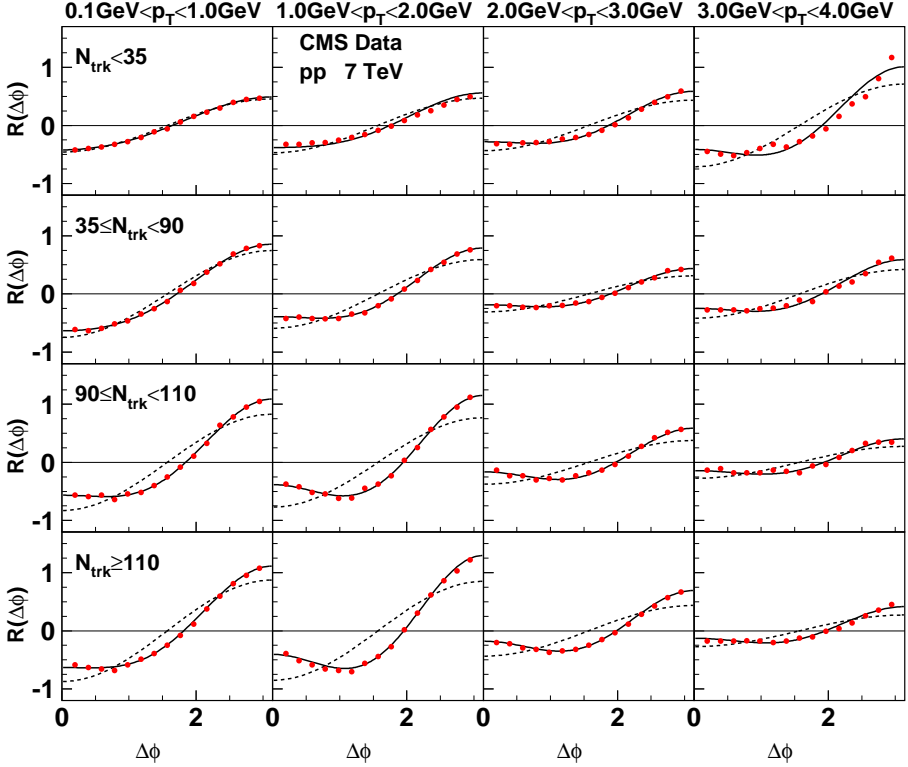}
}
\caption{ The correlation function $R(\Delta \phi)$ as function
 of the relative angle (\ref{eq:Rdef}), for particle pairs with 
pseudorapidity separation $2.0<|\Delta \eta|<4.8$,
  for different multiplicity and transverse momentum
 bins \cite{Khachatryan:2010gv}. The solid lines represent the fit
$-c_1 \cos(\Delta \phi) +c_2 \cos(2 \Delta \phi)$, the dashed lines represent the $-c_1 \cos(\Delta \phi) $ term only.}
\label{fig:rfit}
\end{figure*}

\begin{eqnarray}
f_M(\vec{p}_1,\eta_1,\dots , \vec{p}_M,\eta_M) =&&  \nonumber \\ 
{\delta^2(\vec{p}_1+\dots \vec{p}_M)f(\vec{p}_1,\eta_1)\dots f(\vec{p}_M,\eta_M) } / &&\nonumber \\ 
\int 
\delta^2(\vec{p}_1+\dots \vec{p}_M) &&\nonumber \\
 f(\vec{p}_1,\eta_1)\dots f(\vec{p}_M,\eta_M)
d^2p_1 d\eta_1 \dots d^2p_M d\eta_M  && 
\label{eq:nparticle}
\end{eqnarray}
The  distribution $f$ is
\begin{equation}
f(\vec{p},\eta)= f_{av}(p_\perp,\eta)\left( 1+2 v_2 \cos(2\phi)\right) \ ,
\label{eq:sf}
\end{equation}
and is normalized to one. $M$ is not necessarily the 
total multiplicity in the event, it is rather the number 
of particles created in the microscopic process conserving the momentum.
 $v_2$ is the elliptic flow coefficient, that is
 assumed to have negligible variation over the transverse momentum, 
pseudorapidity or multiplicity bins where the averages are taken.
 The two-particle distribution is
obtained from Eq. (\ref{eq:nparticle}) by integrating 
over the momenta of $M-2$ particles. For large $M$, small $v_2$, 
and to  the order $1/N$,
 one obtains \cite{Borghini:2000cm,Bzdak:2010fd} 
\begin{eqnarray}
f_{2}(\vec{p}_1,\eta_1,\vec{p}_2,\eta_2)&=&f(\vec{p}_1,\eta_1)f(\vec{p}_2,\eta_2)
\left(1+\frac{2}{M}-  \frac{p_1^2}{M  \langle p_\perp^2\rangle_F}\right. \nonumber \\ &
 -& \left.
\frac{p_2^2}{M  \langle p_\perp^2\rangle_F}- \frac{2 
p_1 p_2 \cos(\phi_1-\phi_2)}{M  \langle p_\perp^2\rangle_F}\right) \ ,
\label{eq:f12}
\end{eqnarray}
$\langle \dots \rangle_F$ denotes the average over the full phase space.
The single particle distribution requires one more integration over 
$\vec{p}_2$ and $\eta_2$
\begin{equation}
f_1(\vec{p},\eta)=f(\vec{p},\eta)\left(1+\frac{1}{M}-\frac{p_\perp^2}{M \langle p_\perp^2 \rangle_F}\right) \ . 
\end{equation}
The correlation function $R(\Delta \phi)$ measured by the CMS Collaboration can be expressed as
\begin{eqnarray}
&& R(\Delta \phi)= (<N>-1)  \nonumber \\
&& \left(\frac{\int_B d^2p_1 d\eta_1 d^2p_2 d\eta_2 f_2(\vec{p}_1,\eta_1,\vec{p}_2,\eta_2)\delta(\Delta \phi -\phi_1+\phi_2)}
{\int_B d^2p_1 d\eta_1 d^2p_2 d\eta_2 f_1(\vec{p}_1,\eta_1)f_1(\vec{p}_2,\eta_2)\delta(\Delta \phi -\phi_1+\phi_2)} \right. \nonumber \\
&& \left. -1 \right) \ , 
\label{eq:Rdef}
\end{eqnarray}
where the integration is performed over the considered transverse 
momentum bin and fulfilling the condition on the 
pseudorapidity separation $2.0<|\eta_1-\eta_2|<4.8$, also the numerator and 
the denominator in the above expression are averaged over events in a given 
multiplicity  ($N_{trk}$) class. We 
obtain a simple expression
\begin{equation}
R(\Delta \phi)=-c_1 \cos(\Delta \phi)+c_2 \cos(2\Delta \phi)\ ,
\label{eq:Rfit}
\end{equation}
where the coefficients are
\begin{equation}
c_1=\frac{\langle p_\perp \rangle_B^2 (\langle N \rangle -1)}{\langle p_\perp^2 \rangle_F N_{eff}} \ \ \ , \ \ \frac{1}{N_{eff}}=\langle  \frac{1}{M} \rangle_B
\label{eq:c1}
\end{equation}
and
\begin{equation}
c_2=2 (\langle N \rangle -1) v_2^2 \ .
\label{eq:c2}
\end{equation}
To extract the elliptic flow coefficient the average
 multiplicities $\langle N\rangle$  in each $p_\perp$ bin are  used.
These multiplicities are obtained from the integration of the $p_\perp$
 distribution of particles produced in proton-proton collisions 
at $\sqrt{s}=7$~TeV \cite{Khachatryan:2010us}. For minimum bias events 
the transverse momentum distribution is well described using a Tsallis 
distribution
\begin{equation}
\frac{dN}{d\eta p_\perp d p_\perp}= C \frac{p}{E}
\left(1+\frac{E-m_\pi}{n T}\right)^{-n}
\label{eq:mult}
\end{equation}
 where  the slope parameter $T=0.145$~GeV and $n=6.6$. For our estimate,
 we assume the same distribution in the whole pseudorapidity range, 
with the constant $C$ adjusted to reproduce the total 
multiplicity in the interval $|\eta|<2.4$.
 A noticeable increase of the mean transverse momentum with multiplicity 
is observed
\cite{ATLAS-CONF-2010-024}. To take this effect into account, we assume 
that the slope parameter depends on the multiplicity $N_{trk}$
in the CMS acceptance
 region ($|\eta|<2.4$, $p_\perp>0.4$~GeV)
$T=(0.02+.03 \sqrt{N_{trk}})$~GeV. Such a distribution, when recalculated 
for the ATLAS acceptance, reproduces the increase of the mean
 transverse momentum with multiplicity \cite{ATLAS-CONF-2010-024}.
Using the parameterization (\ref{eq:mult}) we calculate the mean
 multiplicities in different $p_\perp$ bins and multiplicity classes $N_{trk}$.
The events used in the analysis have at least two particles in the chosen 
kinematic range. For small multiplicities we take the  mean multiplicity
 $\langle N \rangle$ only for events fulfilling  the condition that $N\ge 2$. 
The correction is made using 
a Poisson distribution.

The expression (\ref{eq:Rfit}) is used to fit the 
angular correlation function $R(\Delta \phi)$ measured
 in proton-proton collisions at $\sqrt{s}=7$~TeV \cite{Khachatryan:2010gv}. The 
quality of the fit is  excellent.
The  angular correlation has
 two components, the momentum conservation and the elliptic 
azimuthal asymmetry correlations. These effects account for all the angular 
structure observed in the experiment for particles emitted in different 
pseudorapidity regions. 
Some deviations are visible in the 
highest momentum bin, especially in the low multiplicity class.
This may signal the presence of two-particle correlation not of the 
assumed form (\ref{eq:Rfit}). The expansion used in deriving 
momentum conservation effects breaks down for large $p_\perp$ and small 
$N_{trk}$ ($M$). The coefficient $c_1$ ($\simeq 0.4$-$0.9$), 
denoting the strength of the 
momentum conservation correlations, is of the order one, as expected from the 
formula 
(\ref{eq:c1}).

\begin{figure}[h]
\resizebox{0.35\textwidth}{!}{%
  \includegraphics{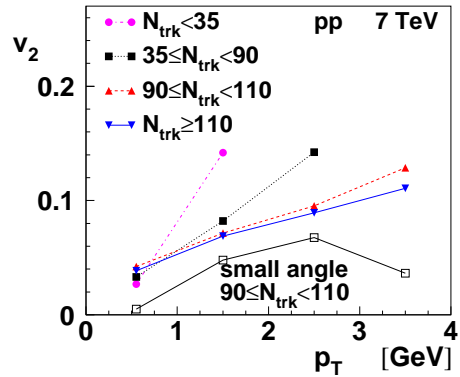}
}
\caption{ The elliptic flow coefficient $v_2$ as function of 
the transverse momentum (Eq. \ref{eq:v2def}), for the 
four multiplicity classes considered. The lower open squares 
represent 
the elliptic flow
coefficient  extracted using the small angle data $\Delta \phi \le  \frac{3 \pi}{8}$, assuming a constant background (Eq. \ref{eq:smallfit}). }
\label{fig:v2}
\end{figure}

The extracted $c_2$ can be used to calculate the strength of the elliptic flow 
in the particle azimuthal distributions 
\begin{equation}
v_2=\sqrt{\frac{c_2}{2(\langle N \rangle -1)}} \ ,
\label{eq:v2def}
\end{equation} which is 
equivalent to the second order cumulant method
\cite{Borghini:2000sa}.
In Fig \ref{fig:v2} is shown the elliptic flow as function 
of the transverse momentum bin for the four multiplicity classes considered.
The first observation is that the results for low multiplicity events differ
from the others. It may indicate that the correlations
 in  low multiplicity events
are of a different origin than in high multiplicity classes. For the high multiplicity events, 
 the value of the elliptic flow increases from $0.03$, in the lowest,
  to  $0.1$, in the highest $p_\perp$  bin. 
This value is too high to be interpreted  as entirely due  to collective
 effects. The strong increase with $p_\perp$, and the large absolute value 
indicate that at least part of the effect is due to the non-flow
 correlations that cannot be separated.
Hydrodynamic model estimates for the elliptic flow coefficient in  
 proton-proton collisions 
are of the order $0.04$-$0.06$, if
 viscosity effects reducing the anisotropy are taken into
 account \cite{Cunqueiro:2008uu,Avsar:2010rf}. 

We have followed the same procedure, using a fit with the first and 
second harmonics,  for the PYTHIA simulated distributions
 \cite{Khachatryan:2010gv}.
The resulting elliptic flow coefficient is of the same order as found in 
the data. This indicates that the non-flow, jet-like correlations, included
 in PYTHIA have both  $\cos(\Delta \phi)$ and $\cos(2\Delta \phi)$ components.
Of course, the elliptic flow component in PYTHIA is not of a collective origin.
On the other hand, the ridge structure is not present in 
the correlation function from 
the Monte Carlo generated events.
Moreover, since the particle pair is separated in pseudorapidity,
 most of the short range correlations, coming from jets or resonance 
decays, are strongly reduced
\cite{Liao:2009ni}. The remaining non-flow correlations are modifying
 the angular correlation function for large relative angles, as seen in  the
 PYTHIA simulations. It indicates that the assumed form of the background 
$-c_1 \cos(\Delta \phi)$ is oversimplified, and that non-flow higher
harmonics are  present in the observed data. The
 small-angle structure can be explained as resulting from the elliptic flow,
 but the extracted values of the $v_2$ coefficient depend on the 
form of the correlation $R(\Delta \phi)$ in the whole range $0$-$\pi$, and
contain
 a large contribution of non-flow origin.
To estimate the value of the elliptic flow related to 
the appearance of the ridge-like structure, we fit a simple formula 
\begin{equation}
R(\Delta\phi)= b+c_2 \cos(\Delta \phi)
\label{eq:smallfit}
\end{equation}
in the range $\Delta \phi \le \frac{3\pi}{8}$. The 
results for the multiplicity class $90\le N_{trk}<110$ 
is shown as the lower points  in Fig. \ref{fig:v2}. 

The momentum resolution of the correlation data 
is not sufficient to
look for a hydrodynamic origin of the effect. 
Future data with more statistics could allow to extract
 a detailed behavior in the soft momentum region $p_\perp<2.0$~GeV. 
The charge independence of the effect found in \cite{Khachatryan:2010gv} 
is consistent with a collective flow origin of the correlations.
The multiplicity dependence of the second harmonic coefficient  
in the correlation function $R(\Delta \phi)$
($c_2 \propto \langle N \rangle $)
 indicates that the underlying correlations ($v_2$) are of a non-flow origin.
 On the contrary, the approximate multiplicity independence of the coefficient 
$c_1$ shows that the underlying correlations decrease linearly
 with the size of the system, which suggests  a non-flow origin of the effect.

Let us comment on the first harmonic term in the correlations function 
$R(\Delta \phi)$. 
The formula (\ref{eq:c2}) allows to estimate the value of effective 
multiplicity $N_{eff}$  of the process for which 
the momentum conservation occurs. We find values between $10$, in 
low $p_\perp$ bins, and $80$, in high momentum and high multiplicity bins. 
Such values are small compared to the multiplicity in the range
 $|\eta|<2.4$, and even more so with respect to the total multiplicity
 in the event. This, and the fact that higher harmonics in the 
angular correlation function are important shows that the observed dominant
correlations have a more complicated form than $\cos(\Delta \phi)$,
 especially in high the $p_\perp$ bins. 

\section{Transverse source velocity}

In this section we analyze the angular correlations assuming a different
 model for the dominant back to back peak. In the first  collision
 between partons a hard momentum transfer occurs. Subsequent particle
 production is determined by the initial transverse momentum exchange.
We use a simple estimate for the angular correlation resulting from such a 
mechanism. Particles separated in pseudorapidity  originate predominantly
from initial 
partons with opposite transverse momenta. Let us assume that in the rest frame of the source connected to a given fragmenting parton, massless particles are emitted isotropically. If the source is moving with velocity $\beta_\perp$ in the direction $\phi=\pi$ the 
final angular distribution takes the form
\begin{equation}
\frac{dN}{d\phi} \propto \frac{\sqrt{1-\beta_\perp^2}}{1+\beta_\perp \cos(\phi)} \ .
\end{equation}
\begin{figure}[h]
\resizebox{0.25\textwidth}{!}{%
  \includegraphics{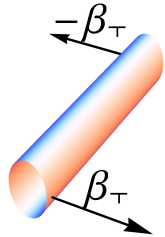}
}
\caption{Schematic picture of the source created in a proton-proton collision. 
At each rapidity it has a transverse 
velocity. The source is azimuthally deformed and  its collective expansion
 would lead to a nonzero elliptic flow.}
\label{fig:rysboost}
\end{figure}
The source from which the other particle in the pair originates, moves on the
 average with velocity $-\beta_\perp$ (Fig. \ref{fig:rysboost}).
 If the transverse momentum and pseudorapidity dependence factorizes from the angular one, we obtain for the angular
 correlation of the pair
\begin{figure*}[ht]
\resizebox{0.75\textwidth}{!}{%
  \includegraphics{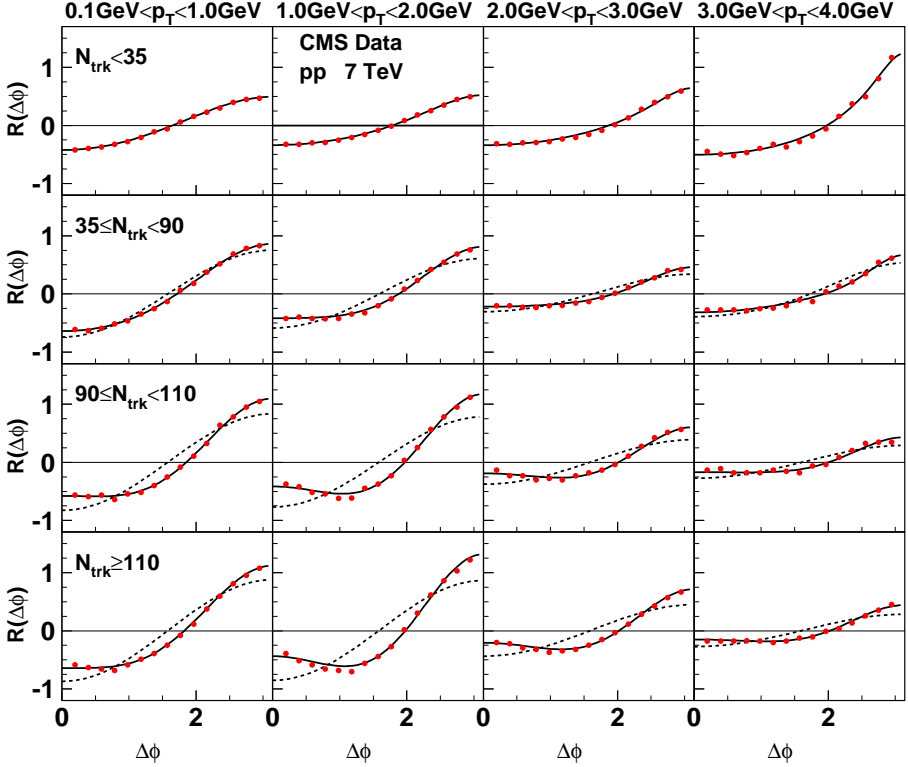}
}
\caption{ The correlation function $R(\Delta \phi)$ as function
 of the relative angle \cite{Khachatryan:2010gv}. 
The solid lines represent the fit
using the boosted source model with elliptic flow  (\ref{eq:rboost}),
 the dashed lines represent correlation for two  boosted isotropic sources (\ref{eq:onlyboost}).}
\label{fig:rboost}
\end{figure*}\begin{equation}
R(\Delta \phi)= (\langle N\rangle -1) \left(\frac{2\sqrt{1-\beta_\perp^2}}{2-\beta_\perp^2+\beta_\perp \cos(\Delta \phi)}-1\right) \ .
\label{eq:onlyboost}
\end{equation}
The above correlation function is peaked for back to back 
emission and contains  the first, second, and higher harmonic components.
In this respect it is different from the simple $-c_1 \cos(\Delta \phi) $ 
expression resulting from the momentum conservation effects.

If the  source is emitting particles in an azimuthally asymmetric
 way in its rest frame 
\begin{equation}
1+2 v_2 \cos(\phi) \ , 
\label{eq:rf}
\end{equation}
 its distribution in the 
center of mass frame is
\begin{eqnarray}
f_\beta(\phi) \propto \sqrt{1-\beta_\perp^2}&& \nonumber \\ \frac{(1+ 2 v_2\left((\beta_\perp+\cos(\phi))^2-(1-\beta_\perp^2)\sin(\phi)\right)}{1+ \beta_\perp\cos(\phi)} \ . &&
\label{eq:ellboosted}
\end{eqnarray}
The correlation function in the relative angle of the pair is
obtained integrating over  the angle of one of  the particles in the pair
\begin{equation}
R(\Delta \phi)=(\langle N\rangle-1)
\left(\int \frac{d \phi }{2\pi}f_\beta(\phi) f_{-\beta}(\phi+\Delta \phi)
-1\right) \ . 
\label{eq:rboost}
\end{equation}
The elliptic asymmetry in the particle distribution 
has the same orientation as  the momentum exchange (source velocity). 
This means, that both
 the hard scattering, that defines the reaction plane, and the 
generated elliptic flow plane have the same orientation. If the generation
 of the 
elliptic flow is a collective expansion from fluctuating initial conditions 
\cite{Bozek:2009dt,CasalderreySolana:2009uk,Ortona:2009yc,Avsar:2010rf}, 
the orientation of the reaction plane from the hard scattering and from 
the elliptic flow is different. In that case, 
the angular distribution in the rest frame
would be $1+2 v_2 \cos(\phi-\Delta \Psi)$. An average
 should should be taken over the 
difference of the two angles $\Delta \Psi$. The results are similar, with 
slightly larger $v_2$, 
as in the scenario using the same orientation of the boost
 and of the elliptic flow (\ref{eq:rf}).

\begin{figure}[hb]
\resizebox{0.35\textwidth}{!}{%
  \includegraphics{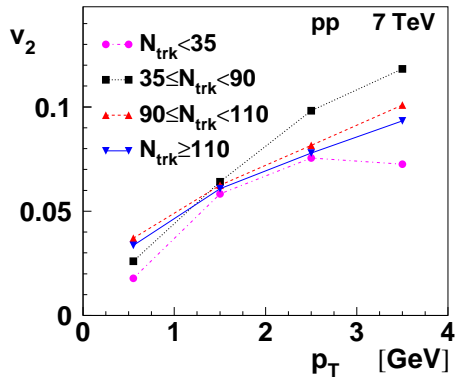}
}
\caption{ The elliptic flow coefficient extracted from
 the CMS data using the  boosted anisotropic source model. }
\label{fig:v2boost}
\end{figure}

\begin{figure}[hb]
\resizebox{0.35\textwidth}{!}{%
  \includegraphics{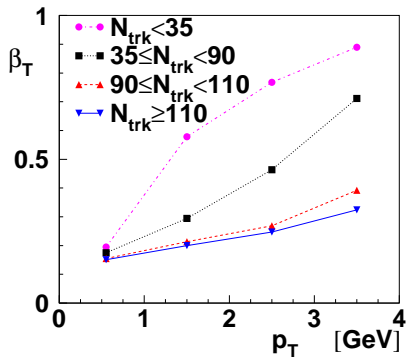}
}
\caption{The  transverse velocity of the 
source extracted from two-particle correlation data. }
\label{fig:boost}
\end{figure}

The formula (\ref{eq:rboost}) is fitted to the measured correlation function
 $R(\Delta \phi)$,
 adjusting two parameters,
 the elliptic flow coefficient $v_2$, and the boost velocity $\beta_\perp$.
The quality of the fit is very good (Fig. \ref{fig:rboost}). In Fig. \ref{fig:v2boost} is shown the 
resulting elliptic flow coefficient $v_2$. It is smaller than in the scenario
 with momentum conservation effects, studied in the previous section.
The extracted elliptic flow increases with the transverse momentum $p_\perp$  
 and depends weakly on the multiplicity.
 In the $p_\perp$
 and multiplicity bins where 
the ridge is observed, no significant rise of $v_2$ is seen. 
The ridge appears because the correlations from the transverse boost of
 the source are smaller. In high multiplicity samples the boost velocity
is much smaller than in the two small multiplicity classes
 (Fig. \ref{fig:boost}). If such transverse boost velocity is 
 present in the 
particle emitting source, it could mimic the  transverse flow  in the
$p_\perp$  spectra of particles.

Our  model of background back to back 
correlations from the transverse boost is very simple. 
We take a single average boost velocity and use 
 massless particles. Going beyond these approximations, would require 
a specific model of the particle emission. The qualitative form of the back to 
back correlation would be similar, but the extracted elliptic flow coefficient 
could depend on the form of the dominant angular correlations.


\section{Conclusions}

We show that the angular correlations between particles emitted 
in proton-proton collisions at $\sqrt{s}=7$~TeV can be decomposed into
 two components.
The  dominant, back to back angular correlations have a kinematic origin.
Modeling them with   momentum conservation effects results in a first harmonic 
component 
in the angular distribution. Another scenario assumes a transverse boost
 of the source, this leads to back to back correlations as well, 
but of a different form.
The remaining correlations, including the small-angle ridge structure,
 are modeled using an elliptic flow contribution. The corresponding 
elliptic flow coefficient $v_2$
 is estimated for the first time in proton-proton collisions. 

The strength of the elliptic flow depends on the assumed model for the
 dominant background, that has to be subtracted. In the momentum 
conservation scenario the elliptic flow coefficient is $0.07$-$0.1$ 
 in the bins where the ridge is observed, whereas in the transverse 
boost estimate it is  $0.06$-$0.08$. The most conservative estimate is obtained
 using a fit of the 
second harmonic only in the small-angle region of the ridge 
(lower points in Fig. \ref{fig:v2}), with the result $v_2=0.04$-$0.06$.
There are several arguments showing that the measured elliptic flow
 contains a collective component. It is approximately multiplicity independent.
The associated structure extends over a long range in pseudorapidity. 
The two particles used to define the second cumulant 
are separated in pseudorapidity, reducing non-flow effects. 
It is tempting to interpret the observed azimuthally asymmetric 
flow as a result of some  strong rescattering or a  hydrodynamic expansion. It 
remains as a challenge to provide an alternative 
 microscopic explanation, as the 
available Monte Carlo generators do not reproduce the observed structures. 
For the event
 generators,  it would require that such azimuthal correlations 
appear early, between most of the particles in the event.
We note that there is no additional  same-side, ridge-like structure in the 
correlation function. All the observed angular 
 correlations are perfectly well accounted for using a combination of 
 kinematic   back to back correlations and of the elliptic flow.

Finally, we stress again the importance of the possible first 
observation  of the elliptic flow in elementary collisions. It would mean that 
the short-lived multiparticle system created in the collision is
 very strongly interacting and  some degree of collectivity  appears.

The author thanks  Adam Bzdak and Wei Li for comments and  discussions 
The work is supported  by the
Polish Ministry of Science and Higher Education 
grant No.  N N202 263438.


\end{document}